\documentclass[conference]{IEEEtran}
   \usepackage[pdftex]{graphicx}
   \graphicspath{{pictures/}}
   \DeclareGraphicsExtensions{.pdf,.jpeg,.png}
\usepackage{todonotes}
\usepackage[cmex10]{amsmath}
\usepackage{textcomp} % z.B. für gerade µ-Zeichen

\usepackage{icomma}
\usepackage{float}

\usepackage{outline}

\usepackage{lipsum}
\usepackage{units}
\usepackage{color}
\usepackage{booktabs}
\usepackage{mdwmath}
\usepackage{mdwtab}
\usepackage{xfrac}
\usepackage[tight,footnotesize]{subfigure}

\usepackage[nolist,nohyperlinks]{acronym}

\usepackage{url}
\usepackage[capitalise]{cleveref}

\begin{document}
\title{A Software-Defined Channel Sounder for Industrial Environments with Fast Time Variance}

\author{\IEEEauthorblockN{Niels Hendrik Fliedner, Dimitri Block, Uwe Meier}
\IEEEauthorblockA{inIT - Institute of Industrial Information Technologies\\OWL University of Applied Sciences\\
	Lemgo, Germany\\
	Email: \{niels.fliedner, dimitri.block, uwe.meier\}@hs-owl.de }
}


%Telephone: (800) 555--1212, Fax: (888) 555--1212}


\maketitle

\begin{abstract}
% Give short view on subject
Novel industrial wireless applications require wide-band, real-time channel characterization due to complex multi-path propagation. 
Rapid machine motion leads to fast time variance of the channel's reflective behavior, which must be captured for radio channel characterization. 
Additionally, inhomogeneous radio channels demand highly flexible measurements. 
Existing approaches for radio channel measurements either lack flexibility or wide-band, real-time performance with fast time variance. 

% Explain novelty of publication in short
In this paper, we propose a correlative channel sounding approach utilizing a software-defined architecture.
The approach enables real-time, wide-band measurements with fast time variance immune to active interference.
The desired performance is validated with a demanding industrial application example.
\end{abstract}


% \fi
\IEEEpeerreviewmaketitle

\begin{acronym}[Bash]
% Communication
\acro{ACI}{adjacent channel interference}
\acro{CTI}{cross-technology interference}
\acro{PDU}{protocol data unit}
\acro{WCS}{wireless communication system}
\acro{SWT}{standardized wireless technology} \acrodefplural{SWT}{standardized wireless technologies}
\acro{QoS}{quality of service}
\acro{UWB}{ultra wide-band}
\acro{VSG}{vector signal generator}
\acro{RTSA}{real-time spectrum analyzer}
\acro{ISI}{inter-symbol interference}

% OSI layers
\acro{PHY}{physical layer}
\acro{MAC}{medium access control}
\acro{DSSS}{direct-sequence spread spectrum}
\acro{CSMA/CA}{carrier-sense multiple access with collision avoidance}
\acro{OFDM}{orthogonal frequency-division multiplex}
\acro{PSK}{phase-shift keying}

% Wireless technologies
\acro{WLAN}{wireless local area network}
\acro{WSAN-FA}{wireless sensor and actuator network
for factory automation}
\acro{HART}{highway addressable remote transducer protocol}
\acro{LR-WPAN}{low-rate wireless personal area network}
\acro{BLE}{Bluetooth Low Energy}
\acro{PNO}{Profibus Nutzerorganisation}
\acro{IWLAN}{industrial WLAN}
\acro{LTE}{3GPP long-term evolution}

% Industry
\acro{BA}{building automation}
\acro{FA}{factory automation}
\acro{PA}{process automation}
\acro{IoT}{internet of things}

% Radio
\acro{ISM}{industrial, scientific and medical}
\acro{ESD}{energy spectral density}
\acro{PSD}{power spectral density}
\acro{ESD}{energy spectral density}
\acro{RF}{radio frequency}
\acro{RSA}{real time spectrum analyzer}
\acro{SNR}{signal-to-noise ratio}
\acro{VSG}{vector signal generator}
\acro{MR}{medium request}
\acro{SIR}{signal-to-interference ratio}
\acro{SINR}{signal-to-interference-plus-noise ratio}
\acro{PAPR}{peak-to-average-power ratio}
\acro{MIMO}{multiple-input and multiple-output}
\acro{DC}{direct current}
\acro{LO}{local oscillator}
\acro{CW}{continuous-wave}
\acro{FMCW}{frequency-modulated CW}
\acro{PDP}{power delay profile}
\acro{PLE}{path-loss exponent}
\acro{AWGN}{additive white Gaussian noise}
\acro{USRP}{universal software radio peripheral}

% GNU Radio
\acro{SDR}{software-defined radio}
\acro{OOT}{out-of-tree}
\acro{DSP}{digital signal processing}
\acro{HDR}{hardware-defined radio}

% Correlation/sounding
\acro{PN}{pseudo-noise}
\acro{MLS}{maximum length sequence}
\acro{FZC}{Frank Zadoff Chu}
\acro{PACF}{periodic auto-correlation function}
\acro{ACF}{auto-correlation function}
\acro{CCF}{cross-correlation function}
\acro{PCCF}{periodic cross-correlation function}

% Error correction
\acro{TOSM}{through-open-short-match}

% Sight conditions
\acro{LOS}{line-of-sight}
\acro{NLOS}{non-line-of-sight}
\acro{OLOS}{obstructed line-of-sight}

% Signal processing
\acro{FFT}{fast Fourier transform}
\acro{IQ}{in-phase and quadrature}
\acro{STFT}{short-term Fourier transform}

% Hardware
\acro{FPGA}{field-programmable gate array}
\acro{LFSR}{linear-feedback shift register}
\acro{AGC}{automatic gain control}

% Architecture
\acro{GNSS}{global navigation satellite system}
\acro{PMT}{polymorphic type}
\acro{OS}{operating system}
\acro{SBC}{single-board computer}

\end{acronym}

\section{Introduction}\label{ch:intro}
%% no \IEEEPARstart
%
%% You must have at least 2 lines in the paragraph with the drop letter
%% (should never be an issue)
%
The recent increase in performance requirements for devices in Industry 4.0 wireless applications demands the development of new \acp{WCS} \cite{2014.FrotzscherWetzker}. 
Such developments require channel characterization in areas previously not fully investigated \cite{2016.HolfeldWieruch}.

%Wireless industrial environments behave very differently compared to indoor office or general outdoor environments.
%\cite{2013.StenumgaardChilo}.
Compared to office and outdoor environments, industrial environments behave drastically worse.
%Multi-path effects are dominant in FA environments. Doppler effect in multi-path environments requires channel characterization \cite{2009.TangWang}.
%Highly absorbent environments: 90\%  of the total received energy within 10 ns delay
%\cite{2013.StenumgaardChilo}.
They are ranging from highly absorbent to highly reflective environments \cite{2013.StenumgaardChilo}.
%-- with the majority of the total received energy falling in the first \unit[10]{ns} delay --
Reflections cause multi-path propagation, which increase frequency variance in \ac{FA} environments \cite{2009.TangWang}.
%Inter-symbol interference due to radio channel time dispersion
%\cite{2016.HolfeldWieruch}.
Multi-path propagation leads to radio channel time dispersion causing \ac{ISI} in \ac{WCS} \cite{2016.HolfeldWieruch}. 
%Latency requirements for Industry 4.0 (and thus for measurements) often stated as \unit[$<$1]{ms}
%\cite{2016.HolfeldWieruch}.
%But, emerging time-critical closed-loop Industry 4.0 applications demand low latency below \unit[1]{ms} \cite{2014.FrotzscherWetzker,2016.HolfeldWieruch}, realized with fast communication signals.
Characterization of channels with multi-path propagation requires wide-band radio channel measurements.

Further, non-stationary multi-path channel conditions also impact \ac{FA} environments \ac{WCS} \cite{2009.TangWang}. 
Especially rapid machine motion causes fast time variance of the channel's reflective behavior.
This challenges time-critical closed-loop Industry 4.0 applications with latencies below \unit[1]{ms} \cite{2014.FrotzscherWetzker,2016.HolfeldWieruch}.
In order to address those fast changes, the radio channel measurement demands to be fast. 

%Generally more complex multi-path propagation in industrial environments. Large varieties in interference levels and multi-path propagation between locations within a single industrial plant
%\cite{2013.StenumgaardChilo}.
Additionally, industrial environments are inhomogeneous, which ranges from highly reflective, complex machine cells to uniform storage areas and transportation paths.
Such inhomogeneity leads to high spatial diversity within \ac{FA} environments \cite{2013.StenumgaardChilo}.
Spatial diversity requires detailed spatial channel characterization, and therefore dense spatial channel measurements. 
Thus, for \ac{FA} environments the measurement methodology has to be flexible, scalable and ideally embedded in a low-cost hardware. 

%TODO: Related Work
Few publications have shown implementation of real-time channel sounding with correlative sequences. 
%TODO: is \cite{2009.GahadzaKim} even real-time?
%\cite{2009.GahadzaKim} achieve \unit[4]{MHz} measurement bandwidth utilizing \acp{SDR} with the GNU Radio framework, which makes this approach unfit for industrial radio channels.
\cite{2016.ElofssonSeimar} use \acp{SDR} with National Instruments' LabVIEW to program the correlation function directly into the \ac{FPGA} for enhanced performance.
Any changes to the measurement setup have to be applied separate from the measurement via \ac{FPGA} programming, which limits the flexibility. 
The hardware-defined HIRATE channel sounder introduced by \cite{2013.KeusgenKortke} and utilized by \cite{2016.HolfeldWieruch} shows even better performance, but conversely is more limited to changes during measurement than \cite{2016.ElofssonSeimar}.
To the best of our knowledge, so far no publication implements a software-defined real-time architecture for industrial radio channel measurements. 

\begin{figure}[H]
	\centering
	\includegraphics[width=0.9\linewidth]{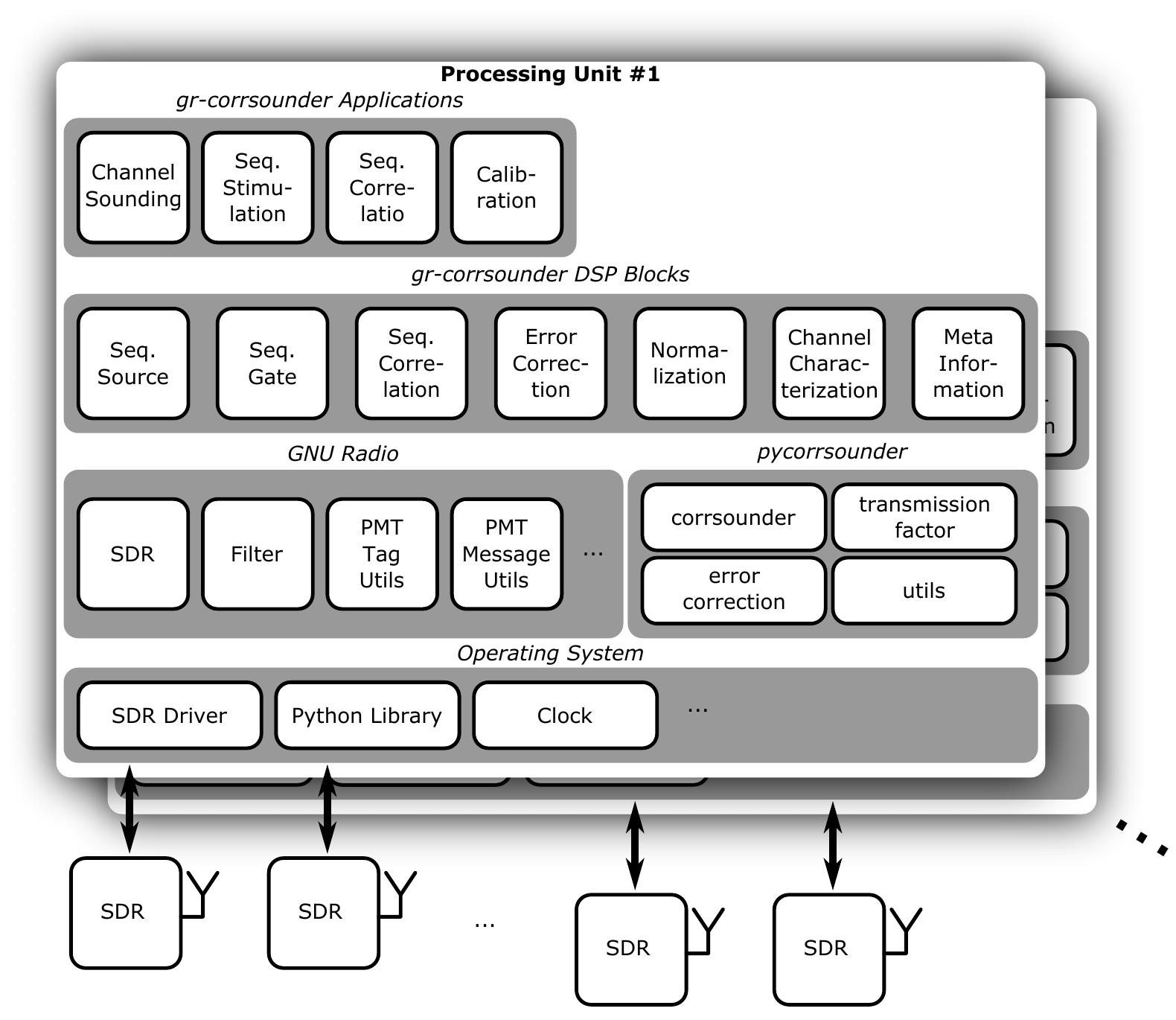}
	\caption{Flexible and scalable architecture of the software-defined wireless channel sounding approach \texttt{corrsounder} for industrial real-time channel characterization}
	\label{fig:gr-corrsounder-architecture}
\end{figure}

In this paper we propose a novel software-defined wireless channel sounding approach named \texttt{corrsounder} for industrial real-time channel characterization. 
In general, channel sounding techniques enable wide-band channel measurements, and therefore enable multi-path propagation characterization.
While channel sounding suffers from limited accuracy, its periodic sequence-based stimulation enable the characterization of fast varying environments.
Additionally, the software-defined realization enables flexible and scalable implementation with off-the-shelf \ac{RF} front ends as shown in \cref{fig:gr-corrsounder-architecture}.  
Furthermore, the proposed real-time capability allows for embedding the measurement system into live processing on demand. 

This paper is structured as follows: \cref{ch:industrialsounding} provides an overview of current solutions for industrial wireless channel sounding approaches, \cref{sec:ProposedModel} describes the the mathematical basics of the utilized correlative channel sounding, \cref{ch:system} details the proposed software architecture of \texttt{corrsounder}, \cref{ch:performance} validates the desired performance with an application example, and finally a conclusion is drawn in \cref{ch:conclusion}. 

\section{Industrial Wireless Channel Sounding}\label{ch:industrialsounding}
% Vermessung industrieller Funkkanäle: 3 Methoden, 2 ausschließen (aber auch Vorteile)
\begin{figure*}[tb]
	\centering
	\includegraphics[width=0.9\linewidth]{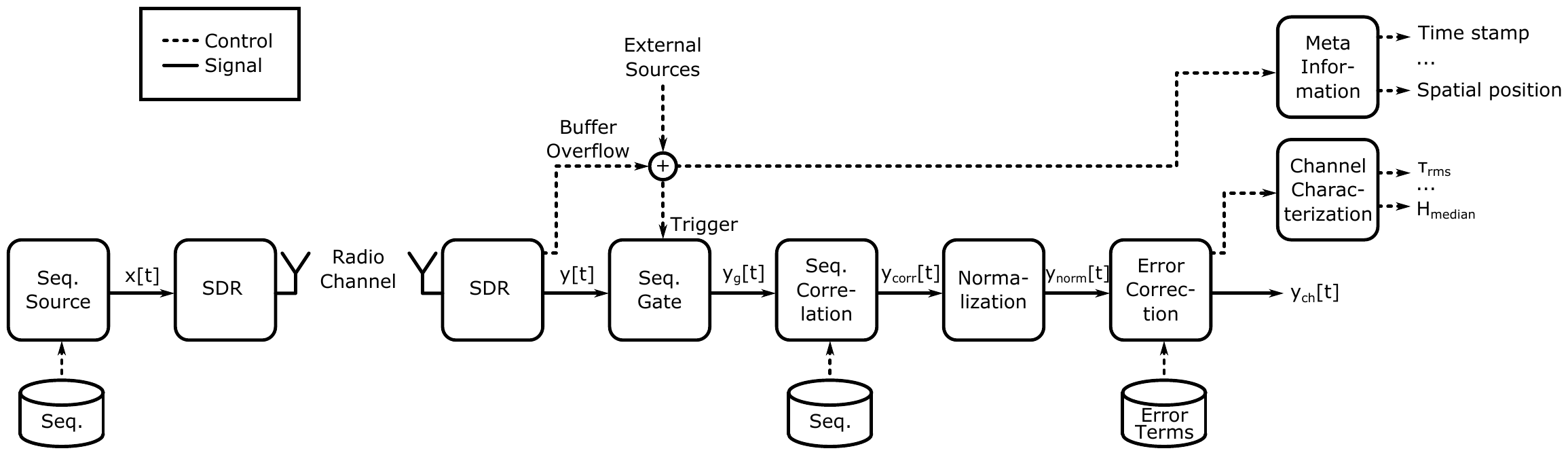}
	\caption{Signal and control flow graph of the sequence-based channel sounding approach}
	\label{fig:gr-corrsounder-signalflow}
\end{figure*}

For channel measurements there exist three methods of channel stimulation:
\ac{CW}, impulse-based and sequence-based stimulation. 
But \ac{CW}- and impulse-based channel characterizations depend on emission measurements.
Hence, they are sensitive to interference.
Therefore, using these methods it is not possible to measure only the frequency or impulse response of an active industrial environment.

Sequence-based stimulations overcome the time limitation and take advantage of the repetitive structure. 
Thereby, various sequences exist for stimulation. 
For reception, sequence-based stimulation require correlation methods, which directly return the scaled impulse response.
These correlation methods suppress the impact of any active interference dissimilar to the utilized sequence.
In contrast to \ac{CW}- and impulse-based stimulations, these methods enable the segregated measurement of the frequency or impulse response of an industrial environment with additional transmitters, which is called an active environment.

Generally speaking, correlative sequences can be differentiated into binary and non-binary ones. 
Typical binary sequences are \ac{PN} sequences.
In general, binary sequences suffer from an inconstant spectral envelope, and therefore linear distortion. 
They are commonly realized in the form of \acp{MLS}, which offer the highest sequence length $N_\text{seq}$ at a certain \ac{LFSR} length $l$, with $N_\text{seq} = 2^l - 1$. 
The \ac{PACF} of \acp{MLS} have a non-zero out-of-phase value $v_\text{oop} = -1$ \cite{2000.Martin}, and thus add a systematic error to the correlation, which limits the dynamic range to $D_\text{r} = N_\text{seq} - |v_\text{oop}|$ \cite{2000.Martin}.
Hence, they are not considered perfect sequences. 
\acp{MLS} provide a constant magnitude, and hence a \ac{PAPR} of one. 
This lowers the requirements on the \ac{AGC} of \ac{RF} frontends, and therefore \acp{SDR}.

For non-binary sequences, there are perfect ones which provide the desired property of a zero out-of-phase \ac{PACF} value and therefore leads to the full dynamic range of $D_\text{r} = N_\text{seq}$.
An optimal choice for a non-binary perfect sequence is called \ac{FZC} sequence \cite{1979.Sarwate}. 
They also provide a constant magnitude, and hence a \ac{PAPR} of one \cite{2011.CaoZhao}. 
Additionally, the spectral envelope of \ac{FZC} sequences is constant without linear distortion. 
Furthermore, these sequences are invulnerable to carrier frequency offset \cite{2011.CaoZhao}, which allows more flexible radio channel measurements in complex scenarios, since no further frequency synchronizing is needed. 
Additionally, multiple \ac{FZC} same-length sequences can be generated, which enables \ac{MIMO} channel measurement \cite{2012.HwangTsai} and multi-user time synchronization, e.g. in \ac{LTE} communication \cite{2015.ETSITS136211}. 

\section{System Description}\label{sec:ProposedModel}

The sequence-based sounding signal flow is illustrated in \cref{fig:gr-corrsounder-signalflow}.
% STIMULATION 
%infinite sequence repetition

For the time-discrete channel stimulation $x[t]$, a signal source outputs infinite repetitions of a chosen correlative sequence $x_\text{seq}[t]$: 
\begin{align}\label{eq:x1}
	x[t] &= \sum\limits_{i=-\infty}^{\infty} x_\text{seq}[t-i \, T_\text{seq}]
\end{align}
with the duration $T_\text{seq} = T_\text{s} \cdot N_\text{seq}$, with the sampling time $T_\text{s}$ and the length $N_\text{seq}$ .
Then, the resulting repetitive signal is transmitted with a \ac{SDR}.

%%% SDR = IQ mod
The transmitting and receiving \acp{SDR} involve \ac{IQ} modulation and demodulation, respectivly.
Additionally, they perform frequency translation.
Hence, the \acp{SDR} handle the low-pass equivalent of a band-pass signal.
So, a band-pass radio channel is stimulated and therfore also evaluated.

% RADIO CHANNEL
%TODO: channel impact. Could also contain frequency offset, DC bias etc., but not discussed here. For frequency offset immunity, see \cite{2011.CaoZhao}. h_\text{ftt} is the deconvolution of h_\text{cable}.
The radio channel influences the transmitted signal:
\begin{align}\label{eq:y1}
y[t] &= x[t] ~\textasteriskcentered~ h[t] + n
\end{align}
with the equivalent low-pass impulse response $h[t]$ and \ac{AWGN} $n$.
%TODO: The former is the impulse response to be ascertained for the channel characterization, and the latter is a cause of incorrect measurement which needs to be compensated with calibration and adjustment prior to the characterization. 
It is important to note, that $h[t]$ contains the impact of the radio channel $h_\text{ch}[t]$ and also the linear distortion caused by utilized antenna equipment such as cables $h_\text{cable}[t]$:
\begin{align}\label{eq:h1}
h[t] &= h_\text{ch}[t] ~\textasteriskcentered~ h_\text{cable}[t]
\end{align}

%TODO: Trigger disrupts continuous computation. Gating guarantees infinite sequence repetition at output. Maybe add dependency on j instead of i to point out repetition and gate-caused cuts.
% SEQUENCE GATE
The following signal processing block, called sequence gate, ensures the repetitive structure of the received signal $y[t]$ for correlation later-on.
The repetitions may be disrupted by trigger events such as reception buffer overflow or from other external sources e.g. \ac{GNSS} localization events.
The sequence gate mutes disrupted sequence durations and passes only full ones.
Its output signal can be expressed as:
\begin{equation}
y_\text{g}[t] = \sum\limits_{j=-\infty}^{\infty} y[t - k_j \, T_\text{seq}] \cdot w_{\text{rect}}\left({(t - k_j \, T_\text{seq})}/{T_\text{seq}}\right)
\end{equation}
with the strictly monotonous growing index $k_j > k_{j-1}$ and the rectangular windowing function $w_{\text{rect}}(t/T)$, which is one for the time interval $\left[0, T\right)$ and otherwise zero.
With no trigger events, $y_\text{g}[t]$ is equal to $y[t]$.

% CORRELATION
Then, the gated signal $y_\text{g}[t]$ is correlated with the chosen stimulation sequence $x_\text{seq}[t]$, which can be expressed as conjugate complex time-reversed convolution:
\begin{align}\label{eq:ycorr1}
	y_\text{corr}[t] &= y_\text{g}[t] ~\textasteriskcentered~ x_\text{seq}^*[-(t-T_\text{seq})]
\end{align}

%TODO: ???? Because the fully correlated output will only start after one sequence duration $T_\text{seq}$, the output is delayed in the time domain. 
%TODO: take assumption from before and apply superposition principle
With no trigger events, $y_\text{corr}[t]$ can be separated into a noise-impact term $y_\text{corr,n}[t]$ and a signal term $y_\text{corr,s}[t]$:
\begin{align}\label{eq:ycorr2}
	y_\text{corr}[t]=\,&(x[t] ~\textasteriskcentered~ h[t]) ~\textasteriskcentered~ x_\text{seq}^*[-(t-T_\text{seq})]\notag\\ 
	&+n ~\textasteriskcentered~ x_\text{seq}^*[-(t-T_\text{seq})]\\
	=\,&y_\text{corr,s}[t] + y_\text{corr,n}[t]
\end{align}
% applied associative property of convolution. 
With good \ac{AWGN} cross-correlation properties of the chosen stimulation sequence as discussed in \cref{ch:industrialsounding}, the impact of $y_\text{corr,n}[t]$ can be neglected.
Then, the associative property of convolution is applied:
\begin{align}\label{eq:ycorrs1}
	y_\text{corr}[t] &= (\,\underbrace{x[t] ~\textasteriskcentered~ x_\text{seq}^*[-(t-T_\text{seq})]}_\text{\ac{PACF}}\,) ~\textasteriskcentered~ h[t]
\end{align}
%TODO: PACF property, perfect sequences: v_\text{oop}=0
The \ac{PACF} property of correlative sequences with constant out-of-phase \ac{PACF} value $v_\text{oop}$ results in:
\begin{align}\label{eq:ycorrs2}
	y_\text{corr}[t] &=\Big( v_\text{oop} + (N_\text{seq} - v_\text{oop})\sum\limits_{i=-\infty}^{\infty} \delta[ t - t_i ] \Big) \textasteriskcentered h[t]
\end{align}
with the measurement time substitution $t_i = (i+1) T_\text{seq} - T_\text{s}$.
%In case of perfect sequences as described in \cref{subsec:CorrelativeSequences}, $v_\text{oop}=0$ always applies. 

%TODO: Sifting property of Dirac function
%TODO: Result: time-shifted and value-scaled impulse response series eq(10) plus a bias error eq(11)
Next, the distributive property of convolution and the sifting property of the Dirac function are applied.
The result is a time-shifted and value-scaled impulse response series.
\begin{align}\label{eq:ycorrs3}
	y_\text{corr}[t] =\,& (N_\text{seq} - v_\text{oop}) \sum\limits_{i=-\infty}^{\infty} h[ t - t_i ]\\ 
	&+ v_\text{oop} \,\textasteriskcentered\, h[t]
\end{align}
Hence, a non-zero $v_\text{oop}$ adds a bias error $v_\text{oop} \,\textasteriskcentered\, h[t]$, which is the case for non-perfect sequences.

%TODO: Noise analysis: longer sequence --> less noise after convolution, see out-of-phase property of cross correlation
In order to return the desired impulse response $h[t]$, the signal $y_\text{corr}[t]$ needs to be scaled according to the number of samples per sequence $N_\text{seq}$.
\begin{align}\label{eq:ynorm1}
y_\text{norm}[t] &= \frac{1}{N_\text{seq}}\cdot y_\text{corr}[t]
\end{align}
This also shows, that the bias error $v_\text{oop} \,\textasteriskcentered\, h[t]$ gets reduced with increased sequence length $N_\text{seq}$.

%TODO: deconvolution with the error terms
The forward transmission tracking error terms are used to correct the linear distortion caused by $h_\text{cable}[t]$, which is still contained within $y_\text{norm}[t]$.
This correction is executed by convolution with $h_\text{ftt}[t]$.
\begin{align}\label{eq:ych1}
	y_\text{ch}[t] &= y_\text{norm}[t]~\textasteriskcentered~h_\text{ftt}[t] \\
	&\approx \sum\limits_{i=-\infty}^{\infty} h_\text{ch}[t - t_i] = \sum\limits_{i=-\infty}^{\infty} h_\text{ch}[\tau, t_i]
\end{align}
with the finely granulated impulse response time $\tau$.

Thus, the proposed measurement method returns the sought-for channel impulse response series.

%%%\section{Related Work}\label{ch:related}
%%%\input{ch-related}

\section{Software Architecture}\label{ch:system}
The channel sounding approach \texttt{corrsounder} is realized based on a modular software architecture.
In the following sub-sections the software architecture shown in \cref{fig:gr-corrsounder-architecture} is discussed.

\subsection{Distributed Processing}
Channel sounding with sequence-based stimulations can be splitted into the transmission part for stimulation and the reception part for correlation.
These are independent from each other as shown in \cref{fig:gr-corrsounder-signalflow}, as long as an identical sequence is provided.
Hence, channel sounding stimulation and correlation may be separated, and the associated \acp{SDR} can be connected to multiple processing units such as embedded host computers.
Each processing unit has to be connected to at least one \ac{SDR}, performing either the stimulation or correlation processing.
With multiple \acp{SDR} it may perform any arbitrary combination of the stimulation or correlation processing.
Therefore, the software architecture approaches a distributed processing.
It also enables integration into existing utilized wireless infrastructure, and therefore increases flexibility.

\subsection{Operating System}
%%% Linux
The \texttt{corrsounder} expects a Linux-based \acp{OS} for the processing units.
Hence, general-purpose Ubuntu \ac{OS} for a high-power x86 hardware architecture, but also low-power embedded ARM-based Linux derivates for \ac{SBC} can be addressed.
%%% RT Drivers
The \acp{OS} have to provide real-time hardware bindings to the \acp{SDR} such as device drivers.
In case of additional trigger sources, the bindings such as hardware clocks and \ac{GNSS} receivers also have to be provided by the \ac{OS} .
Additionally, the \ac{OS} has to provide the Python runtime and the computation library \texttt{numpy} for signal processing.

\subsection{Kernel Module}
%%% Kernel module
The kernel module \texttt{pycorrsounder} contains basic routines for channel sounding operation as well as for evaluation.
It only requires the Python related libraries.
Thus, the dependencies of \texttt{corrsounder} for pre- and post-processing are limited up to the kernel module, which increases the flexibility.
%%% Routines
\texttt{pycorrsounder} contains the following routines:

\paragraph{Sequence Generation and Correlation}
Generation routines are provided for binary \ac{MLS} and non-binary \ac{FZC} sequences.
Additional sequence generation routines can be added later on.
The module also contains routines for sequence correlation: \ac{ACF}, \ac{PACF}, \ac{CCF}, and \ac{PCCF}.
The \ac{ACF} can be used for evaluation of a finite signal, while \ac{PACF} is the counterpart for assumed infinite repetition of the finite signal.
The \ac{CCF} and \ac{PCCF} can be used accordingly for evaluation of two non-equal signals, such as a transmitted sequence and the corresponding received channel response. %or for \ac{MIMO} channel characterization evaluation of two sequences.

\paragraph{Error Correction and Filtering}

%%% systematic linear error correction
% Ref: Agilent Application Note AN 1287-3 "Applying Error Correction to Network Analyzer Measurements", 2002
For two-port systematic linear error correction with two \ac{SDR} antenna ports, a through adjustment routine is provided.
It requires the error term for forward transmission tracking.
The error term can be acquired by directly connecting the \ac{SDR} antenna ports.
Through adjustment is easy to perform, removes the main frequency response error and can be utilized when medium accuracy is required.
%%% Filtering
Various other distortion effects can be suppressed by the provided discrete filtering routines.
A removal routine for peaks and a down-sampling routine are provided.
%%% Fade-out
The removal routine can suppress spectral and temporal peaks such as the \ac{DC} bias.
The bias is caused by the \ac{LO} \ac{RF} leakage of a receiving \acs{SDR} \cite{2005.SvitekRaman}.
It results in a constant value offset and therefore a spectral peak around \unit[0]{Hz}.
Hence, the removal routine fades out such spectral frequency range and interpolates the values afterwards.
%TODO: Find reference for fade-out routine. Der alte Wiltron macht das auch!
%%% Down-sampling
The down-sampling routine is utilized to suppress high-frequency attenuation by extracting the low-frequency part.
%TODO: due to \ac{SDR} pre-filtering???
The routine determines a cutoff frequency at which the \ac{PSD} falls below a threshold relative to the maximum, e.g. \unit[-3]{dB} for the edge frequency.

\paragraph{Channel Characterization}

Routines are provided for channel characterization.
Thereby, impulse response time $\tau$ and measurement time $t$ evaluation routines are provided for single-sequence and inter-sequence characterization, respectively.
The former characterize the frequency and the impulse response at a certain measurement time $t$.
Thereby, the \ac{PDP} $E[|h(\tau)|^2]$, the mean delay $\bar{\tau}$ and the delay spread $\tau_\text{rms}$ can be characterized.
For the frequency response, the routines characterize the \ac{PSD} $E[|H(f)|^2]$ and statistical metrics like percentile \unit[10]{\%} $H_{10}$, \unit[90]{\%} $H_{90}$, \unit[50]{\%} $H_\text{median}$ of the magnitude $|H(f)|$ and the coherence bandwidth $B_C$.
Additionally, arbitrary distributions can be estimated like the Rician distribution for \ac{LOS} conditions.
Furthermore, the \ac{PLE} can be estimated based on the transmission and reception antenna gain.
Measurement time characterization routines include for example the Doppler spread $h[\tau, f_d]$ and the coherence time $T_C$.

\subsection{GNU Radio}

In case the processing unit has to perform real-time processing, the well-known \ac{SDR} framework GNU Radio is required.
GNU Radio is a library of \ac{DSP} blocks, \ac{SDR} bindings and application routines with several command-line and graphical user interfaces.
It is utilized in several simulation studies as well as for \ac{DSP} application deployment.
For the signal and control flow \cref{fig:gr-corrsounder-signalflow} different GNU Radio approaches are applied.
For the signal flow, stream-based items of complex \ac{IQ} samples have to be utilized.
GNU Radio provides in-tree \ac{DSP} blocks for stream processing such as \ac{SDR} source and sink, 
convolutional and DFT-based filter blocks.
In contrast, the control flow can also be event-based for triggers.
Therefore, GNU Radio provides in-tree blocks for event-based processing with \ac{PMT} tags and \ac{PMT} messages.
\ac{PMT} tags are synchronous events which annotate signal streams.
\ac{PMT} messages are asynchronous events which are processed concurrently and independent of the signal streams.

\subsection{\texttt{corrsounder} DSP Blocks}

The \texttt{corrsounder} \ac{DSP} blocks are provided within the \ac{OOT} GNU Radio module \texttt{gr-corrsounder}.
It requires a GNU Radio instance and the routines of the kernel module \texttt{pycorrsounder}.
\texttt{gr-corrsounder} includes the \ac{DSP} blocks sequence source, sequence gate, sequence correlation, error correction, channel characterization, and the meta information.
These blocks are already discussed in \cref{sec:ProposedModel}.

\subsection{\texttt{corrsounder} DSP Applications}

The module \texttt{gr-corrsounder} also contains the \texttt{corrsounder} \ac{DSP} applications.
These are executable flow graphs of connected \ac{DSP} blocks.
Mainly, four applications are provided.
The first application is called \textit{Channel Sounding}.
It represents the channel sounding signal and control flow shown in \cref{fig:gr-corrsounder-signalflow}.
It combines two applications: \textit{Sequence Stimulation} and \textit{Sequence Correlation}.
\textit{Sequence Stimulation} is an application for the stimulation part with the transmission-related processing blocks.
\textit{Sequence Stimulation} is the counterpart for reception-related part.
The final application is called \textit{Calibration}.
It is a derivate of the \textit{Channel Sounding} application without the error correction blocks.
It has to be utilized for calibration to determine the error terms.

\subsection{Source Code Availability}
The source code of the channel sounder \texttt{corrsounder} is online available via the URL
\url{https://github.com/inIT-HF/gr-corrsounder}.
It contains the module \texttt{gr-corrsounder} with the \ac{DSP} blocks and applications as well as the kernel module \texttt{pycorrsounder}.

%%%\section{Measurement Setup}\label{ch:meassetup}
%%%\input{ch-meassetup}

\section{Application Example}\label{ch:performance}
An exemplary measurement demonstrates the performance of our proposed channel sounder architecture.
The radio channel consists of a robot arm performing a pick-and-place routine inside a separated cell. 
Two antennas are placed in opposite corners with a distance of \unit[3.8]{m} and a height of \unit[1.5]{m}.
The robot arm obstructs the \ac{LOS} periodically during motion.

We utilize distributed processing, with a \ac{VSG} for stimulation and a host computer connected to an \ac{SDR} for correlation. 
The stimulation sequence is generated by the kernel module \texttt{pycorrsounder}.
For correlation, the \texttt{corrsounder} \ac{DSP} application \textit{Sequence Correlation} is used.

\begin{table}[htp]
	\caption{Measurement Setup and Parameters}
	\label{tab:MeasurementInstrumentsParameters}
	\centering
	\begin{tabular}{ll}
		\toprule
		Unit & Value \\
		\midrule
		\ac{VSG} & Rohde\& Schwarz SMBV100A\\
		Antennas & WiMo 18720.3H \unit[5]{GHz} \\
		Antenna gain & \unit[5]{dBi} \\
%		Antenna directivity & omnidirectional \\
		Robot arm & Reis RV16 \\
		\ac{SDR} & Ettus Research USRP X300 \\
%		\ac{SDR} interface & PCI-Express x4 \\
%		Host computer CPU &  Intel XEON E5-1660 v3\\
%		Host computer RAM & \unit[16]{GB}\\
		Host computer \ac{OS} & Ubuntu Linux 14.04\\
%		GNU Radio version & 3.7.11\\
		\midrule
		Center frequency & \unit[5.80]{GHz} \\
		Sampling rate & \unit[100]{MSps} \\
		Sequence type & \ac{FZC} \\
		Sequence length & \unit[1024]{samples} \\
		\ac{FZC} parameter & 7 \\
		Measurement \unit[3]{dB} bandwidth & \unit[93.07]{MHz} \\
		Calibration routine & Through \\
		\ac{DC} bias suppression bandwidth & \unit[781]{kHz} \\
		Measurement duration & \unit[12]{s} \\
		\bottomrule
	\end{tabular}
\end{table}

With the utilized hardware listed in \cref{tab:MeasurementInstrumentsParameters}, the proposed channel sounder architecture enables real-time measurement without buffer overflows with the mentioned performance. 

\begin{figure}[H]
	\centering
	\includegraphics*[width=0.8\linewidth]{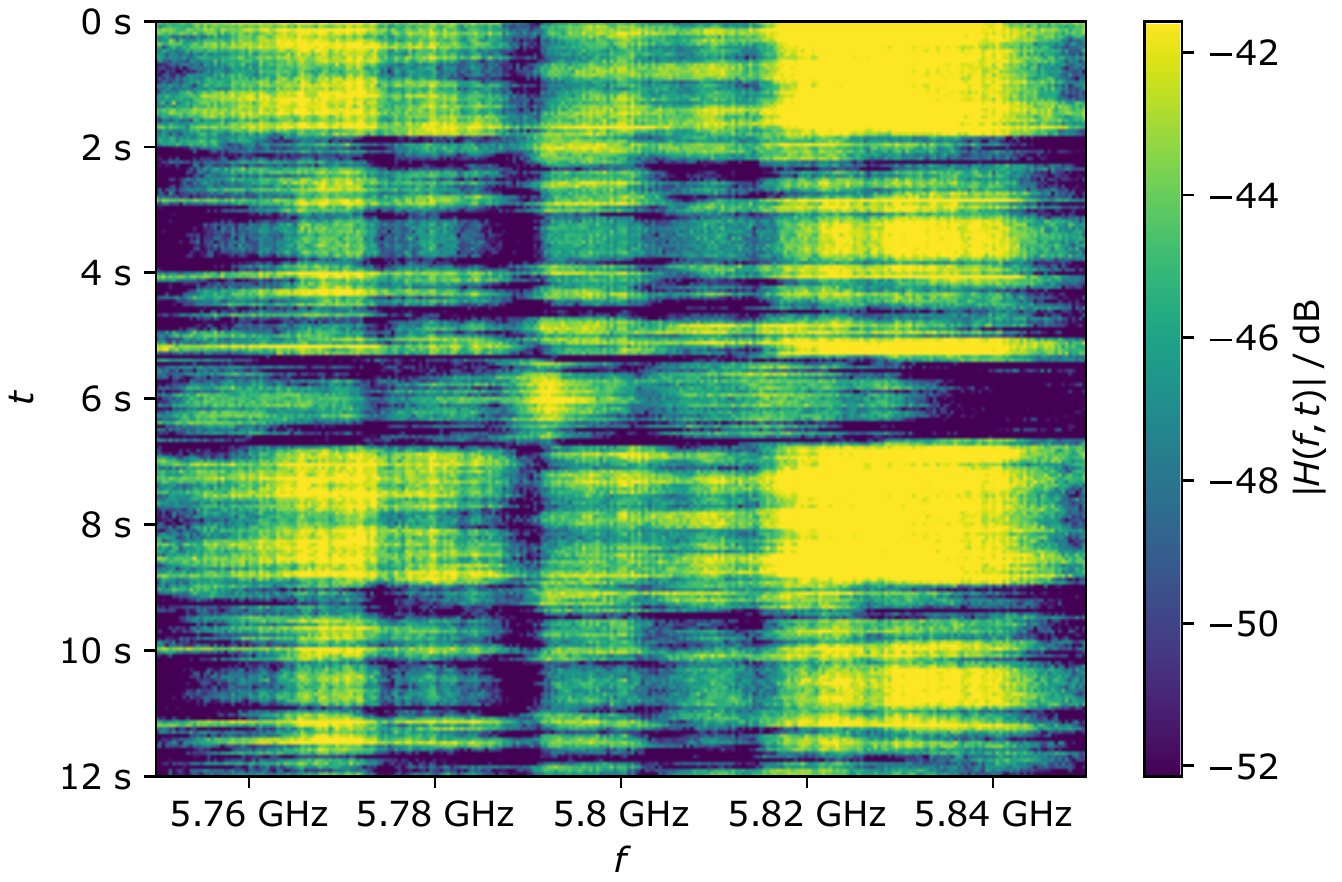}
	\caption[TEST]{Frequency response}
	\label{fig:frequency_response}
\end{figure}

%Additionally, \unit[12]{s} of raw measurement data are captured using the RAM disc, as shown in \cref{fig:frequency_response}. 
The motion of the robot arm causes a frequency response pattern with a repetition time of approximately \unit[7.5]{s}, as shown in \cref{fig:frequency_response}.
In contrast, the measurement time resolution of \unit[10.24]{\textmu s} falls far below the time variance of the fast motion.
Therefore, our proposed approach surpasses the fast time variance requirements for industrial applications with even higher demands.

\begin{figure}[H]
	\centering
	\includegraphics*[width=0.8\linewidth]{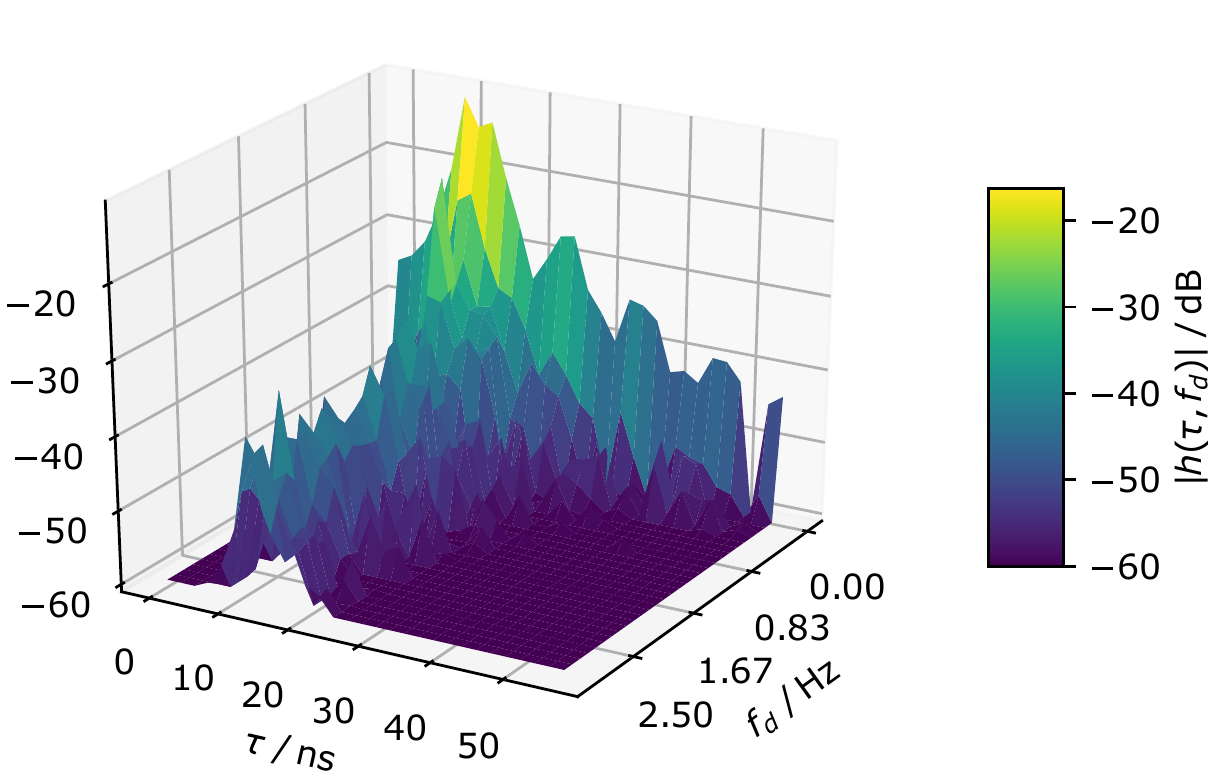}
	\caption[TEST]{Doppler plot}
	\label{fig:doppler_plot}
\end{figure}

The Doppler plot in \cref{fig:doppler_plot} shows the impulse response and indicates the time variance of the multi-path propagation.
The plot resolves distinct multi-path components in the finely granulated impulse response time resolution of \unit[10.75]{ns}. 
Furthermore, the long-term measurement enables Doppler spread resolution of \unit[0.083]{Hz} for time-sensitive characterization of the multi-path propagation.
Additionally, the maximum resolvable Doppler spread of \unit[48.8]{kHz} corresponds to a speed of \unit[2526]{m/s}.

The application example measurement results in a dynamic range of \unit[45.1]{dB}.
This dynamic range enables a measurement distance of \unit[695.87]{m} under ideal propagation conditions.

%Since the robot arm causes comparably slow time variance in the radio channel compared to industrial environments with rapid machine motion, much faster time variance can occur in the Doppler plot for other applications. 
%%Rapid motion happens especially in highly automated production environments. 
%This fast variance in the radio channel can only be mapped using fast time-variant channel measurements. 

%During measurement, the application example offers real-time correlation and plotting, which can be utilized for external real-time applications, e.g. radio channel estimation for coexistence management. 

The application example validates, that the proposed channel sounder architectures is suited for wide-band, real-time radio channel measurements with fast time variance. 

\section{Conclusion}\label{ch:conclusion}
%\todo[inline]{Niels und Dimi}
Novel industrial wireless applications require wide-band, real-time channel characterization due to complex multi-path propagation. 
Fast time variance of the channel's reflective behavior must be captured for radio channel characterization. 
Additionally, inhomogeneous radio channels demand highly flexible measurement measurements. 
Existing approaches for radio channel measurements either lack flexibility or wide-band, real-time performance with fast time variance. 

The proposed approach utilizes a software-defined architecture for flexibility and scalability.
Furthermore, the approach uses correlative channel sounding for real-time, wide-band measurements immune to active interference.
Additionally, the approach enables distributed processing, and therefore it is favorable for embedded implementation.

In an application example, the distributed processing property is validated for characterization of a radio channel obstructed by robot arm motion. 
Hence, the proposed correlative channel sounding approach is suited for industrial wide-band, real-time channel characterization with fast time variance.


\section*{Acknowledgment}
Parts of this work are funded by the German BMBF (project HiFlecs, 16KIS0266).

\bibliographystyle{IEEEtran}
\bibliography{references}
%\begin{thebibliography}{1}


%\end{thebibliography}

\end{document}